# Topological design and synthesis of high-spin aza-triangulenes without Jahn-Teller distortions

James Lawrence[1†], Yuanyuan He[2†], Haipeng Wei[1†], Jie Su[1], Shaotang Song[1], Alina Wania Rodrigues[3], Daniel Miravet[3], Pawel Hawrylak[3,4], Jianwei Zhao[2], Jishan Wu*[1], Jiong Lu*[1,4]

[1]Department of Chemistry, National University of Singapore, 117543, Singapore
[2]College of Material and Textile Engineering, Key Laboratory of Yarn Materials Forming and Composite Processing Technology of Zhejiang Province, Jiaxing University, Jiaxing 314001, Zhejiang, PR China
[3]Department of Physics, University of Ottawa, Ottawa, Ontario, K1N 6N5, Canada
[4]Institute for Functional Intelligent Materials, National University of Singapore, 117544, Singapore

*†* These authors contributed equally.

Email: chmwuj@nus.edu.sg; chmluj@nus.edu.sg

**Abstract:**

The atomic doping of open-shell nanographenes enables the precise tuning of their electronic and magnetic state, which is crucial for their promising potential applications in optoelectronics and spintronics. Among this intriguing class of molecules, triangulenes stand out with their size-dependent electronic properties and spin states, which can also be influenced by the presence of dopant atoms and functional groups. However, the occurrence of Jahn-Teller distortions in such systems can have a crucial impact on their total spin and requires further theoretical and experimental investigation. In this study, we examine the nitrogen-doped aza-triangulene series *via* a combination of density functional theory and on-surface synthesis. We identify a general trend in the calculated spin states of aza-[n]triangulenes of various sizes, separating them into two symmetry classes, one of which features molecules that are predicted to undergo Jahn-Teller distortions that reduce their symmetry and thus their total spin. We link this behavior to the location of the central nitrogen atom relative to the two underlying carbon sublattices of the molecules. Consequently, our findings reveal that centrally-doped aza-triangulenes have one less radical than their undoped counterparts, irrespective of their predicted symmetry. We follow this by demonstrating the on-surface synthesis of π-extended aza-[5]triangulene, a large member of the higher symmetry class without Jahn-Teller distortions, *via* a simple one-step annealing process on Cu(111) and Au(111). Using scanning probe microscopy and spectroscopy combined with theoretical calculations, we prove that the molecule is positively charged on the Au(111) substrate, with a high-spin quintet state of S = 2, the same total spin as undoped neutral [5]triangulene. Our study uncovers the correlation between dopant position and the radical nature of high spin nanographenes, providing a novel strategy for the design and development of these nanographenes for various applications.

**TOC graphic:**

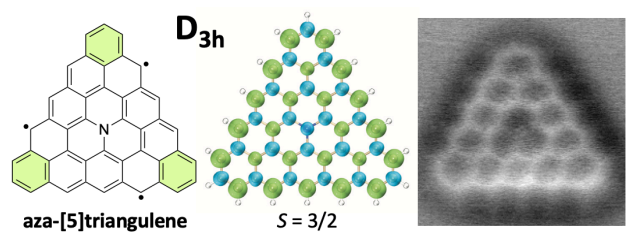

Triangulenes are a class of fascinating open-shell non-Kekulé nanographenes that have sparked extensive studies since the pioneering on-surface synthesis (OSS) of [3]triangulene in 2017 by Pavliček et al.[1–5] Their non-trivial π-magnetism stems from an inherent imbalance in their π-electron network that can be elucidated by applying Lieb's theorem[6] or Ovchinnikov's rule[7] to their bipartite lattice structures. Notably, one sublattice harbors a greater number of carbon sites compared to the other sublattice, thereby often predicting the number of unpaired electrons in each molecule. Previous works have demonstrated a direct correlation between the size of triangulenes and the extent of this sublattice imbalance, consequently influencing their ground spin states.[8–11] The open-shell nature of triangulene and its various derivatives has typically been proven in such OSS studies through the observation of Kondo peaks in d$I$/d$V$ spectra,[12–14] singly occupied/unoccupied molecular orbitals[1,8,9] and collective spin excitations,[15] forming a comparison with accompanying theoretical spin-polarized calculations.

As they possess reactive zigzag edges[16] and unpaired electrons, triangulenes are notoriously difficult to synthesize and isolate in solution. This is the main reason that unsubstituted triangulene was first observed by using OSS under ultra-high vacuum conditions (UHV),[1] a remarkable 64 years after its first conception by Clar and Stewart in 1953.[17] While some progress has been made in the solution-phase synthesis of chemically-protected triangulenes,[18–23] the majority of the experimental research efforts have primarily taken place in the OSS field, alongside theoretical studies.[11,24] The combined use of techniques such as scanning tunnelling microscopy (STM), scanning tunnelling spectroscopy (STS) and non-contact atomic force microscopy (nc-AFM) also allows the precise study of the electronic properties of individual triangulene molecules adsorbed on surfaces following their synthesis. This has enabled researchers to examine a variety of triangulene-based structures and their spin states, such as extended triangulenes,[13,25] short chains,[14,26,27] quantum rings[28] and nanostars.[15]

The atomic doping of triangulenes (and nanographenes generally[29–34]) enables the precise tuning of their magnetic state. There are several examples of smaller doped triangulenes in the literature, with heteroatoms either incorporated within their structure[12,18,35,36] or at their zigzag edges.[21] Among these, the largest examples are [4]triangulenes with BO or BN-doped edges.[37–39] Doping triangulenes with heteroatoms alters the number of unpaired electrons originating from the sublattice imbalance. In the case of aza-triangulenes, it is expected that replacing a central carbon (which can be viewed as having a single unpaired electron in its p-orbital) with a nitrogen (with its accompanying lone pair) will deprive one sublattice of an unpaired electron, thereby modifying the total spin of the molecule. In this simplified picture, the radical character of an aza-triangulene would depend on whether the nitrogen atom is located on the same majority sublattice as the zigzag carbons (reduces overall spin by decreasing the lattice imbalance) or the minority sublattice (increases overall spin by increasing the lattice imbalance). However, Wang et al.[12] recently synthesized aza-[3]triangulene on Au(111) and Ag(111) and theoretically predicted that neutral aza-[3]triangulene undergoes Jahn-Teller distortions,[40] lowering its symmetry from $D_{3h}$ to $C_{2v}$ and thus reducing its spin from $S = 3/2$ to $S = 1/2$. This discovery has opened up new possibilities for the rest of the aza-triangulene series. Nevertheless, it still remains elusive whether we can design an alternative class of π-extended aza-triangulenes with both a high total spin and a high symmetry, requiring further experimental and theoretical investigation.

In this study, we first employ density functional theory (DFT) calculations to further enhance the understanding of aza-[n]triangulenes, focusing on larger molecules within the series. By investigating these π-extended structures, we demonstrate that aza-[n]triangulenes can be classified into two distinct groups with varying symmetries, depending on the position of the central nitrogen atom in relation to the underlying carbon sublattices. Notably, we find that the aforementioned Jahn-Teller distortions of the class with $C_{2v}$ symmetry nullifies the effect on total spin that the different sublattice nitrogen positions would have otherwise had, leading to all aza-triangulenes studied being predicted to have one fewer radical than their undoped counterparts, regardless of their calculated symmetry. In order to experimentally validate these findings, we then use a straightforward OSS procedure to fabricate a large aza-[5]triangulene on Cu(111) and Au(111) surfaces. We utilize a suitably designed precursor that only requires surface annealing, without the extra steps of hydrogenation and tip-induced dehydrogenation that were required by Wang et al.[12] Both the atomic structure and electronic properties of aza-[5]triangulene on Au(111) and Cu(111) can be subsequently characterized using bond-resolved scanning probe microscopy and spectroscopy, respectively, corroborated by theoretical calculations.

# Results and discussion:

**Predicting two sets of aza-triangulenes, with and without Jahn-Teller distortions**

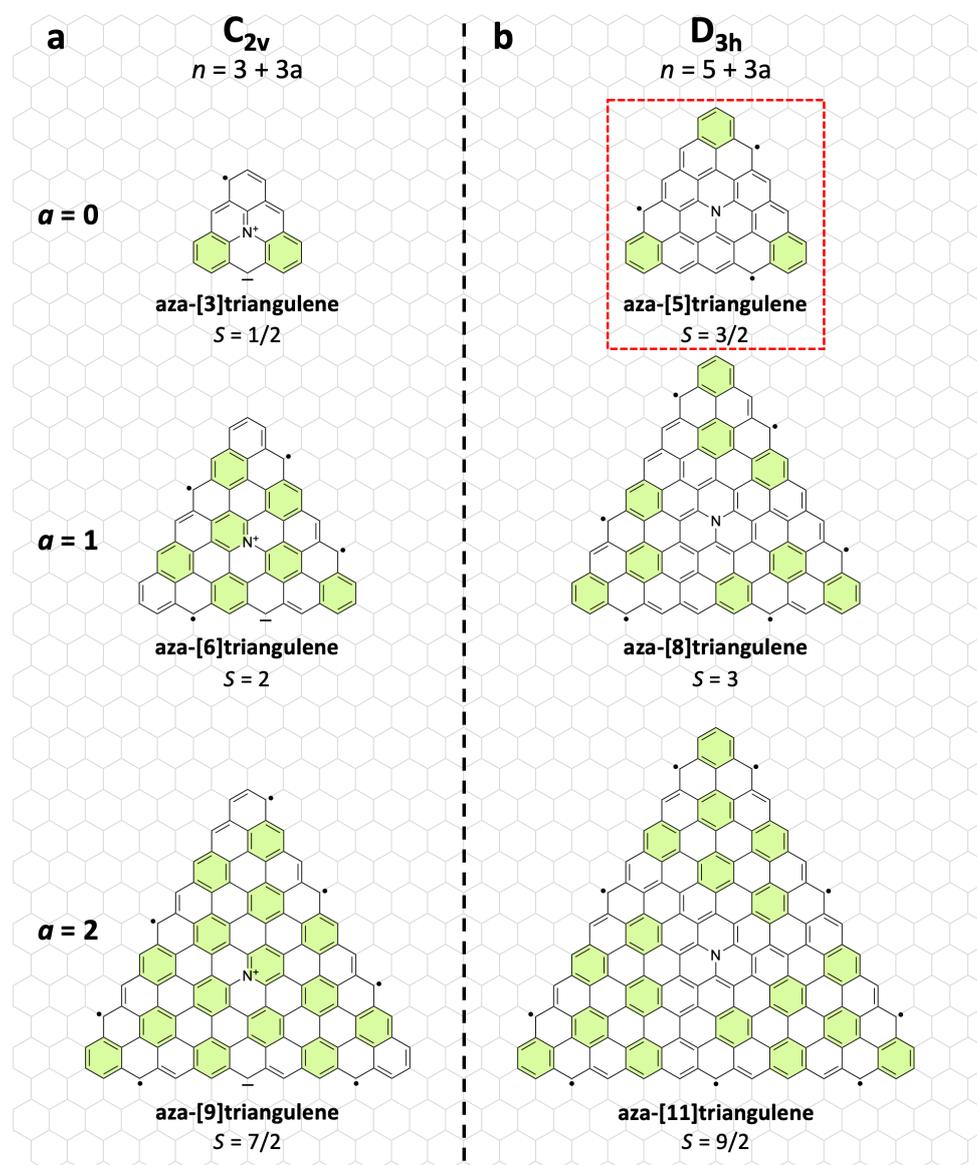

**Figure 1.** The chemical structures of the two different classes of centrosymmetric aza-[n]triangulenes with their calculated lowest energy spin states. Each pair (e.g. [3] and [5]) have been assigned an integer, $a$ ($a$ = 0, 1, 2, …), that is related to the number of zigzag edge sites, $n$. The two classes have a different relation between $a$ and $n$, as shown at the top of the figure. (a) The $C_{2v}$ class of molecules, which undergo Jahn-Teller distortions to reduce their spin and lower their energies. These distortions are represented in the structures by a positively charged nitrogen atom and negatively charged zigzag edge carbon atom. Drawing them in this way also reduces the number of necessary radicals in the chemical structure. (b) The $D_{3h}$ class, which do not distort. Drawing the structures of this class with charged atoms does not affect the resulting number of radicals (refer to fig S7). The Clar sextets of the aza-[n]triangulenes are shaded in green. The main subject of this study, π-extended aza-[5]triangulene, is highlighted in a red box.

We first performed gas phase spin-polarized DFT calculations for the first six molecules in the series of centrosymmetric nitrogen-doped aza-triangulenes: aza-[3], [5], [6], [8], [9] and [11]triangulene. These calculations reveal the lowest energy spin states for each of these structures (figure 1). The calculated energy levels and orbital densities for all of these molecules are presented in figure S2, together with detailed calculations for aza-[5]triangulene in different charge states (figures S3-S5). It becomes immediately apparent from the DFT-optimized geometries of the aza-triangulenes that they can be divided into two distinct sets, depending on their energetically-favored symmetries. These two groups can be categorized according to the relationship between the number of zigzag edge sites ($n$) and an associated integer $a$, where $a$ = 0, 1, 2, and so on. The aza-[3 + 3$a$]triangulenes in figure 1(a) are calculated to have a $C_{2v}$ symmetry, whereas the [5 + 3$a$] set in figure 1(b) are found to be $D_{3h}$. This extends the findings of a previous study for aza-[3]triangulene in which the neutral molecule has a $C_{2v}$ symmetry, changing to $D_{3h}$ upon one-electron oxidation.[12,18] This is associated with Jahn-Teller distortions that reduce the symmetry of the neutral [3 + 3$a$] class to $C_{2v}$ to minimize their energy,[40] with a corresponding reduction in the number of radicals and thus their total spin.

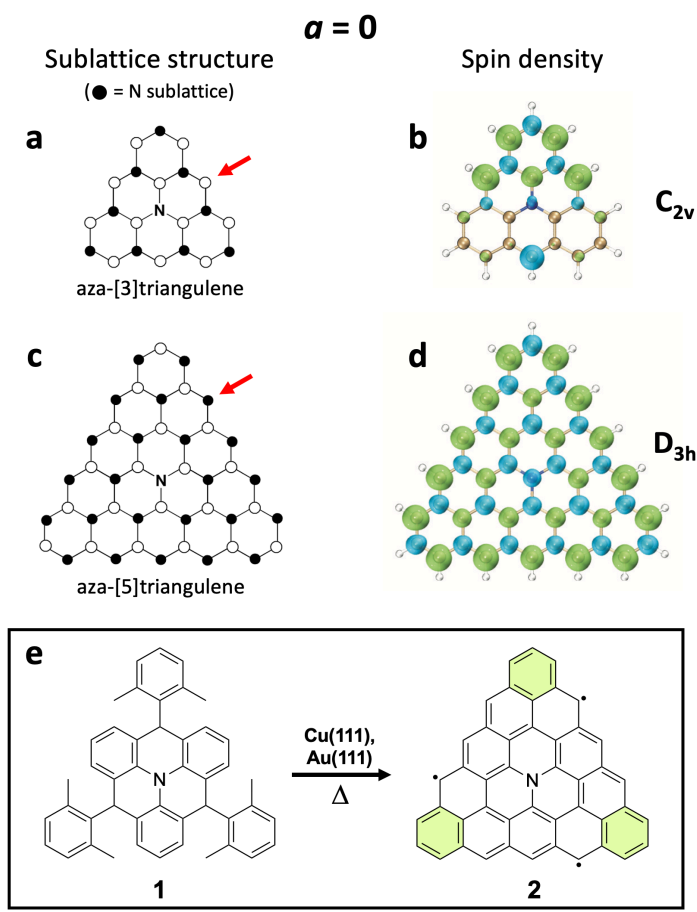

**Figure 2.** (a)-(d) Sublattice structures and spin densities of the $a$ = 0 aza-[n]triangulenes. In the case of the aza-[3]triangulene, the nitrogen atom inhabits a different sublattice to the carbon zigzag edges (indicated by a red arrow), whereas for the π-extended aza-[5]triangulene they inhabit the same sublattice. The different symmetry of each molecule is clear in the accompanying spin densities in (b) and (d). (e) shows the proposed OSS reaction scheme to form π-extended aza-[5]triangulene. The three Clar sextets of the product are shaded in green.

The existence of these two symmetry classes is related to the position of the nitrogen atom in their sublattice structures. To illustrate this, the sublattice structures and calculated spin densities of the two molecules with $a = 0$ are shown in figure 2: aza-[3]triangulene and aza-[5]triangulene. For the $[3 + 3a]$ $C_{2v}$ molecules (figure 1(a) and figure 2(a)/(b)), the central nitrogen atom sits on a different (minority) sublattice to the zigzag edge carbons, whereas for the $[5 + 3a]$ $D_{3h}$ set they occupy the same (majority) sublattice. This is depicted in figure 2(a) and (c), where the nitrogen-hosting sublattice of each molecule is drawn with filled circles and the zigzag edge carbons are indicated with red arrows.

The Jahn-Teller distortions of aza-$[3 + 3a]$triangulenes (and lack thereof for the $[5 + 3a]$ set) can be intuitively chemically rationalized by drawing out resonance structures (figures S6 and S7, table S1), both with and without including charged nitrogen/carbon atoms. The C=N double bond in the charged resonance structures represents the bond length reduction that is found for one of the C-N bonds in the DFT-optimized structures of the $C_{2v}$ set. For this set, the number of radicals is reduced when including charged components in the resonance structure to lower their symmetry, whereas the molecules calculated as $D_{3h}$ retain the same radical character and total spin in both cases. Compared to undoped triangulenes, the effect of these distortions is to change the influence of the nitrogen atom's sublattice position on the total spin of the molecules. The positively-charged central nitrogen atom can now be viewed as contributing a single electron to the minority sublattice that it is located on, with the negatively charged zigzag edge carbon removing an unpaired electron from the majority sublattice. This reduces the lattice imbalance by 2 compared to a resonance structure drawn with only neutral atoms, lowering the total spin and thus the energy of the molecule. For the $[5 + 3a]$ $D_{3h}$ group, charging the nitrogen and the zigzag edge carbon has an overall net zero effect on their spin as they both occupy the same sublattice. As a result of these effects, all of the aza-[n]triangulenes shown here have one fewer radical than their undoped counterparts, despite any different sublattice occupancy by the nitrogen atom.

As an illustrative example: using Lieb's theorem,[6] undoped [5]triangulene is expected to have 4 radicals ($S = 2$) due to its sublattice imbalance, with 25 carbons on sublattice A (which includes the zigzag edges) and 21 carbons on sublattice B. Aza-[5]triangulene is calculated here to possess 3 radicals ($S = 3/2$). This fits with the expectation that the central nitrogen atom (located on sublattice A) removes an unpaired electron from the *majority* sublattice due to the lone pair in its $p_z$ orbital, thus reducing the overall spin of the molecule by 1/2. Likewise, undoped [6]triangulene is predicted to have 5 radicals ($N_A = 33$, $N_B = 28$) with $S = 5/2$. The original simple model would thus expect that the nitrogen atom in aza-[6]triangulene (located on *minority* sublattice B, differently to the zigzag edges) reduces $N_B$ by 1 and thus *increases* the number of radicals to 6 ($S = 3$). However, the calculations here show that aza-[6]triangulene in fact distorts to have $C_{2v}$ symmetry and 4 radicals, one fewer than the undoped molecule – the same difference that aza-[5]triangulene has with [5]triangulene, despite its different symmetry group.

**On-surface synthesis and characterization of aza-[5]triangulene**

To experimentally validate these findings, in the following we demonstrate the synthesis and characterization of aza-[5]triangulene, one representative example of the higher symmetry $D_{3h}$ class, *via* on-surface chemistry on two metallic substrates: Au(111) and Cu(111). In our case, the molecule can be synthesized by simply annealing a precursor molecule on the metal surfaces in a purely bottom-up fashion. The precursor molecule designed here (compound **1**, figure 2(e)) contains a hydrogenated aza-[3]triangulene core with three 2,6-dimethyphenyl groups (for details of its synthesis, refer to the Supporting Information). This precursor is structurally related to a molecule formed in a recent work that synthesized an aza-[3]triangulene in solution,[18] and bears a resemblance to the precursor used in our earlier work for the synthesis of [5]triangulene on surfaces.[9]

In order to form the desired planar product, the first step of the on-surface reaction presumably involves the three $sp^3$ carbons of the core of the precursor undergoing dehydrogenation. Following this, the two methyl groups on each of the outer phenyl rings can then fuse with an aromatic ring in the core *via* a ring-closing reaction, with the loss of three hydrogen atoms in total for each fusion. We have shown previously that a very similar reaction occurs with a closely-related precursor that forms undoped [5]triangulene on the same metal surfaces.[9] Compound **1** was first sublimed under ultra-high vacuum conditions onto room temperature Cu(111) and Au(111) surfaces. Post-annealing these preparations had two different outcomes, depending on the metal substrate used.

Annealing the Cu(111) sample to 250 °C led to a high yield of small objects that were evenly distributed over the surface terraces (figure 3(a)). Closer inspection of these *via* STM revealed their triangular shape (figure 3(b)), and bond-resolved STM (BR-STM, figure 3(c)) and nc-AFM (figure 3(d)) images further elucidated their internal bonding structure, which fits well with the expected aza-[5]triangulene product. The contrast of the darker rings around the nitrogen atom in the BR-STM image is fairly typical for rings containing heteroatoms,[12,41,42] and may be attributed to variations in the tip-molecule interactions near the heteroatom[42] and/or the density of states for the molecule near the Fermi level. Varying the tip-molecule distance during imaging shows how these dark features evolve as the CO tip nears the molecule (figure S8). In the nc-AFM image in figure 3(d), the bonds at the apexes of the molecule appear with a more repulsive (brighter) contrast, indicating that they may be slightly raised above the rest of the molecule. Within the central region there is also a slightly more repulsive contrast on the bonds, similar to calculated nc-AFM images of undoped [5]triangulene.[9] This adsorption conformation also bears some similarity to a previous study of undoped [7]triangulene on Cu(111), in which it was found to have a slightly dome-shaped configuration.[10] This may be related to hybridization with the surface and charge transfer.

As was also the case for undoped [5]triangulene,[9] the reaction yield on Au(111) was much lower than Cu(111) and required a higher activation temperature. Annealing to 310 °C quickly led to a mixture of intra- and inter-molecular products. While most of the objects that can be seen in the STM images (figure 3(e) and (f)) are flat and have thus undergone the required fusion of the methyl groups with the aromatic rings, they are also mostly bound to other molecules. Some partially reacted molecules were also still present, with bright features that correspond to methyl groups that had not yet fused with adjacent rings. Tweaking the annealing temperature slightly (figure S9) clearly demonstrates that there is a significant overlap with the temperatures required for activating the intra- and inter-molecular reactions, as reducing the temperature by only 10 °C led to mostly individual partially reacted molecules being present on the surface.

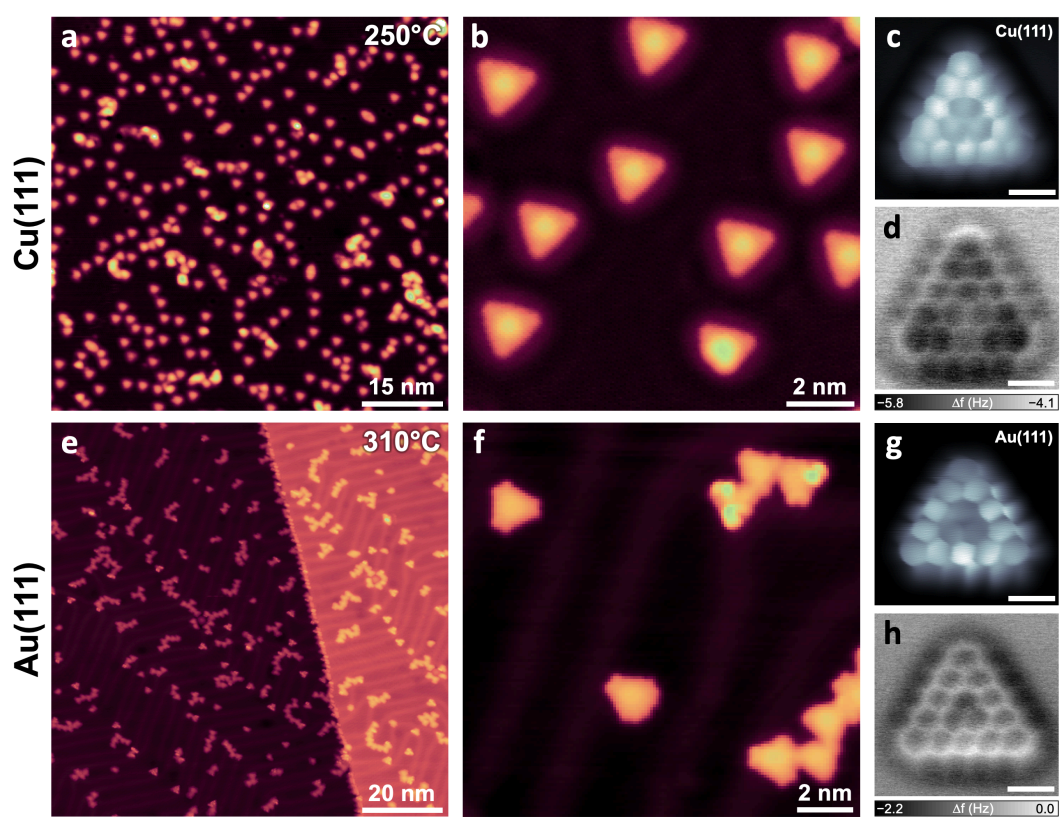

**Figure 3.** (a)-(b) and (e)-(f): STM images recorded after annealing precursor **1** on Cu(111) and Au(111) respectively. For details of the annealing procedures, refer to the methods section. (c) and (g): CO tip BR-STM images of π-extended aza-[5]triangulene on Cu(111) and Au(111), respectively. (d) and (h) constant height frequency shift nc-AFM images of the same aza-[5]triangulene molecules as (c) and (g), recorded with the same CO tip. Scanning parameters: (a) 60 pA, +500 mV (b) 100 pA, +500 mV (c) constant height, +5 mV (d) constant height, 50 pm oscillation amplitude (e) 100 pA, −700 mV (f) 100 pA, −1 V (g) constant height, +5 mV (h) constant height, 50 pm oscillation amplitude.

Despite this, intact aza-[5]triangulene molecules were occasionally found on Au(111), as shown with the BR-STM and nc-AFM images in figure 3(g) and (h). On Au(111), the molecule does not present the same brighter apex/central features in the nc-AFM imaging when compared to Cu(111). This may indicate that the molecule is more planar on Au(111). The nc-AFM image of aza-[5]triangulene on Au(111) fits well with a particle-probe simulation[43,44] for imaging a free-standing planar molecule (figure S10), particularly for the dark central rings around the nitrogen atom. Some molecules were also found with their most reactive zigzag edge carbons passivated with extra hydrogen atoms, as shown in figure S11. These were easily removed by moving the STM tip nearby with typical tunnelling currents (2-300 pA) and higher positive bias voltages (+2 V). Bond-resolved imaging before and after this process shows a good agreement with previous studies of hydrogenated zigzag nanostructures.[1,16,45] As there were a large number of partially reacted molecules on Au(111) after annealing, tip-induced reactions could also be used to form aza-[5]triangulene. This is shown in more detail in figure S12.

DFT-optimized adsorption structures of aza-[5]triangulene on both Au(111) and Cu(111) are shown in figure S13. These calculations predict that the molecule is adsorbed significantly closer to the Cu(111) surface (2.17 Å) than the Au(111) surface (2.64 Å). Furthermore, its conformation on Cu(111) is calculated to be significantly less planar: both the central region of the molecule around the nitrogen and the hydrogens at the apexes are raised relative to their neighbors. This fits well with the aforementioned features observed in nc-AFM images of the molecule on Cu(111) (figure 3(d)). Likewise, the slightly more planar calculated adsorption on Au(111) matches the fairly uniform contrast observed in nc-AFM imaging in figure 3(h).

**Electronic characterization of aza-[5]triangulene**

Magnetic nanographenes typically exhibit a strong electronic interaction with the Cu(111) substrate, which can result in the absence of open-shell character. As Au(111) is generally more inert than Cu(111), it is assumed to have less of an effect on the observed molecular density of states when compared to gas-phase calculations. As such, the results on Au(111) may provide a better basis for a comparison with gas-phase calculations that do not include the effect of the substrate. Therefore, we primarily focus on the electronic characterization of aza-[5]triangulene adsorbed on the Au(111) substrate by recording $dI/dV$ spectra and images. A line of $dI/dV$ spectra, represented by a heat map, is shown in figure 4(a). The spectra were recorded along the edge of an aza-[5]triangulene molecule on Au(111), as indicated in the inset STM image. Two examples of $dI/dV$ spectra (recorded at the two positions displayed in the inset image) are overlaid onto this heatmap, with the peaks of interest highlighted by dashed lines. The onset in the occupied molecular states occurs at approximately −1 V, with peaks marked at −960 mV and −1200 mV. The lowest unoccupied state is positioned at +180 mV, with higher states at around +860 mV and +1560 mV. In comparison to this, $dI/dV$ spectra taken over the molecule on Cu(111) (shown in figure S14) reveal the presence of a similar onset in the occupied states, but with a very different set of states in the unoccupied region. This may be further evidence for a stronger interaction and thus greater level of hybridization with the Cu(111) substrate than for Au(111).

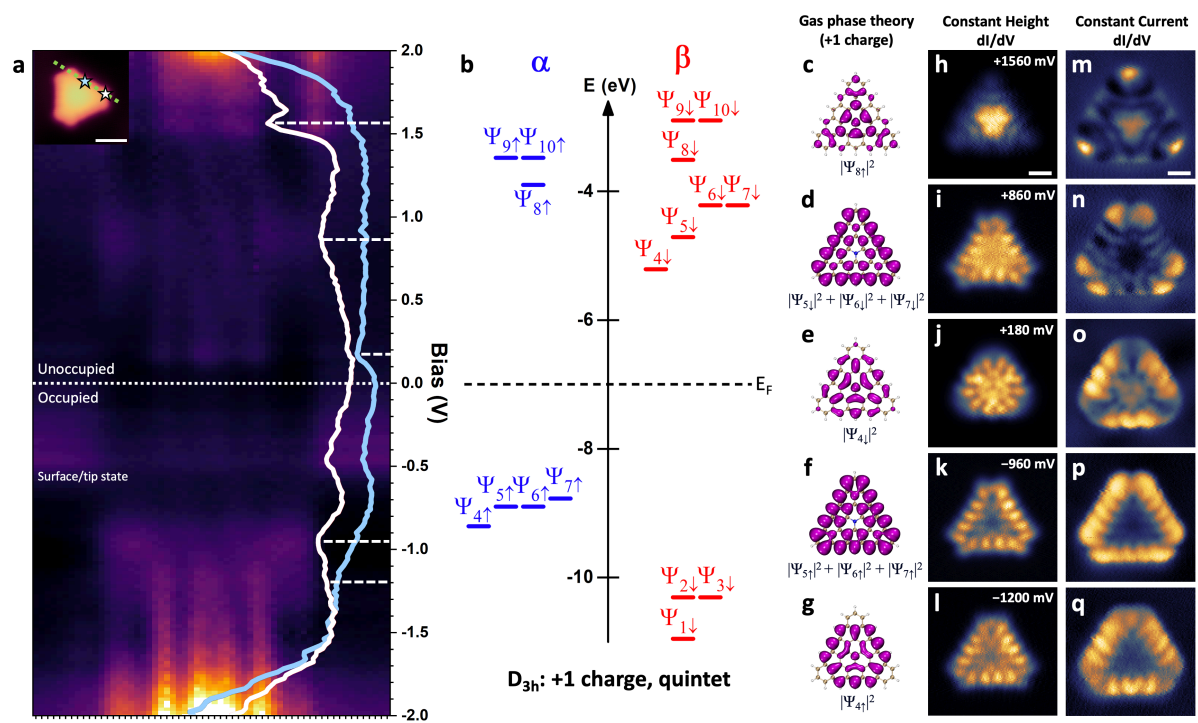

**Figure 4.** The electronic structure characterization of aza-[5]-triangulene on Au(111). (a) A stacked line of d$I$/d$V$ spectra recorded along the edge of an aza-[5]triangulene molecule on Au(111), as indicated in the inset image. Two examples of d$I$/d$V$ spectra that were recorded at different positions on the edge of the molecule are overlaid. The positions at which they were recorded are marked by stars in the inset image. The most important features in the spectra are marked with dashed lines. (b) Spin-polarized DFT calculations for the singly positively charged molecule. Presented to the right of this are the calculated orbital densities (c)-(g) and CO tip d$I$/d$V$ images (both constant current and constant height, (h)-(q)) of an aza-[5]triangulene molecule recorded at voltages corresponding to the main features in the d$I$/d$V$ spectra. Orbitals that are close in energy are shown with a combined density, as the energy resolution in d$I$/d$V$ experiments was most likely not able to differentiate them. The choice of combinations involves some nuance, and several options are discussed in the main text and SI. Scanning parameters: the constant current images in (m)-(q) were recorded at tunneling current setpoints of 2.0 to 2.5 nA. The scale bar of the inset STM image in (a) is 1 nm and the scale bars in (h) and (m) are 500 pm.

Spin-polarized DFT calculations for a singly positively charged aza-[5]triangulene are shown in figure 4(b), with the neutral equivalent presented in figure S3. All wavefunctions and corresponding orbital densities are presented in figures S4 and S5. Alongside the calculated combined frontier orbital densities for the positively charged molecule in figure 4(c)-(g) are matching d$I$/d$V$ images (figure 4(h)-(q)) recorded at the voltages corresponding to the aforementioned peaks in the d$I$/d$V$ spectra. These were recorded with a CO-terminated tip in both constant height and constant current modes. Images of these states show the expected edge-localized nodal patterns that fit well with previously observed states at the zigzag edges of carbon nanostructures.[46] The images at constant height help to show the relative decay into the vacuum of these states at different parts of the molecule, whereas the constant current images highlight the features that may become less obvious due to this decay. They can also generally help to normalize the effects of any slight non-planarity in the conformation of molecules on the resulting imaging. As the structure of the tip can sometimes play a role in the appearance of the underlying molecule, a comparison between d$I$/d$V$ images recorded with both a CO tip and a metallic tip is presented in figure S15. The main features used to identify the states are present in both sets of data.

The molecule is assumed to be positively charged on Au(111) due to the low-lying unoccupied state observed in the d$I$/d$V$ data at approximately +180 mV. For the spin-polarized calculations on the neutral molecule, the highest singly occupied molecular orbital (SOMO) is $\Psi_{4\downarrow}$ (figure 5), which then becomes the lowest singly unoccupied molecular orbital (SUMO) for the positively charged molecule. Comparing the CO tip d$I$/d$V$ images recorded at +180 mV in figure 4(j) and (o) with the calculated orbital density $|\Psi_{4\downarrow}|^2$ shows striking similarities, indicating that the molecule has donated a single electron to the high workfunction Au(111) substrate. Making a comparison between the d$I$/d$V$ images and the other orbitals/orbital combinations requires more nuance, as is shown in more detail in figures S16 and S17.

In particular, a choice is made to combine the orbital densities of $\Psi_{5\downarrow}$, $\Psi_{6\downarrow}$, and $\Psi_{7\downarrow}$ to form a comparison with the images recorded for the broad unoccupied feature observed at +860 mV (figure 4(i) and (n)), despite there being a significant energy gap in the calculations between $\Psi_{5\downarrow}$ and $\Psi_{6\downarrow}$/$\Psi_{7\downarrow}$. This choice is informed by the resemblances of the other orbital densities for $\Psi_{8\uparrow}$ and $\Psi_{4\downarrow}$ to the images recorded for the peaks above (+1560 mV) and below (+180 mV) in voltage in the d$I$/d$V$ data. It should be noted that there is almost always a significant difference between gas-phase DFT calculations and calculations for molecules adsorbed on a metallic surface, so these states may in fact be closer in energy in the adsorbed situation, presumably due to interactions with the metallic substrate and electronic screening.[47]

For the occupied states, two features can be observed in the d$I$/d$V$ line data: an initial onset at around −960 mV with more intensity towards the apexes of the molecule, and a feature beginning at −1200 mV, which has more intensity in the center of the edge with a clear nodal pattern seen in the line spectra. The d$I$/d$V$ signal then generally increases as the voltage becomes more negative. As there are only four singly-occupied orbitals near the highest occupied onset in the DFT calculations (figure 4(b)), it is clear that they can be divided into two groups: the lower energy (more negative voltage) $\Psi_{4\uparrow}$ orbital, with its higher density in the central edge of the molecule, and a combination of $\Psi_{5\uparrow}$, $\Psi_{6\uparrow}$ and $\Psi_{7\uparrow}$ for the higher energy (less negative voltage) feature in the d$I$/d$V$ spectra, with a more even distribution of features along the edges of the molecule and a slightly higher intensity at the apexes. As these states are so close in energy, they cannot be fully isolated from one another due to a lack of energy resolution during d$I$/d$V$ experiments. As such, the d$I$/d$V$ images at these voltages (figure 4(k), (p), (l), (q)) are both 'cross-contaminated' with one-another, with features from the adjacent state/s also being present. This explains why there is still some d$I$/d$V$ signal observed at the apexes of the molecule in the image at −1200 mV, whereas the matching $\Psi_{4\uparrow}$ orbital has no density there.

All of the SPM data serve to confirm the calculated $D_{3h}$ symmetry of the positively charged aza-[5]triangulene molecule, as all of the observed states appear with threefold symmetry. Any slight asymmetries that may be visible in images can be attributed to either the tip or the position of the molecule on the Au(111) surface relative to its herringbone reconstruction, which can have small effects on both the adsorption conformation and the voltage at which features appear in d$I$/d$V$ spectra, due to a slightly varying surface potential.

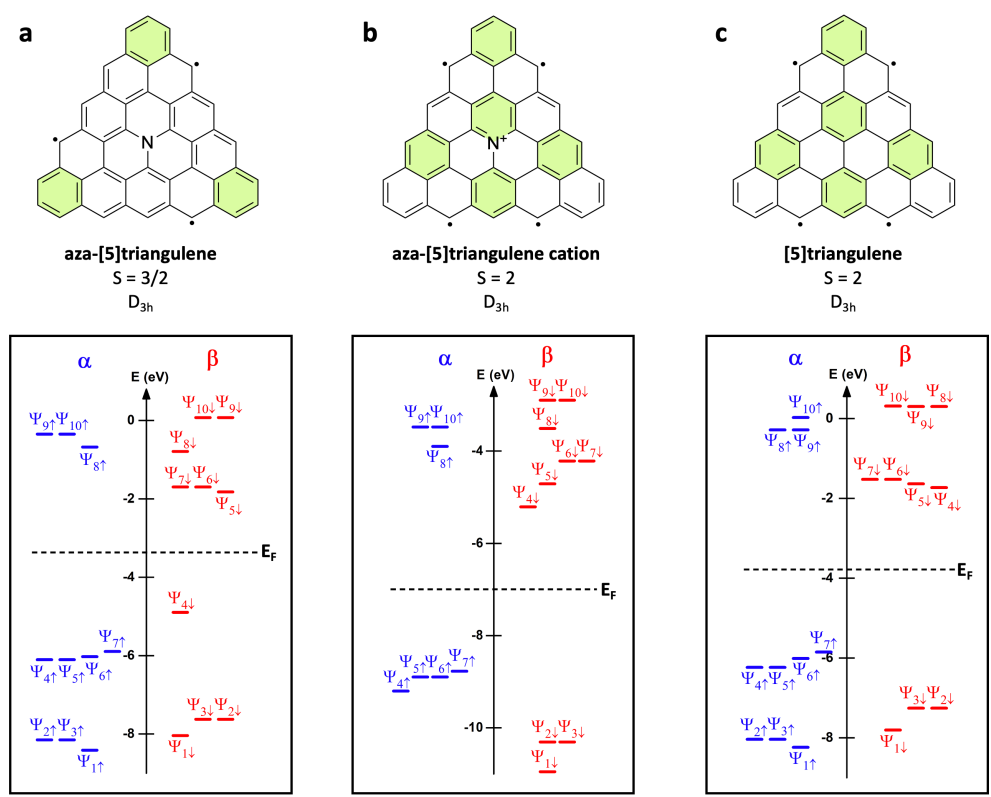

**Figure 5.** (a)-(c) Chemical structure, total spin and symmetry of gas-phase aza-[5]triangulene, positively-charged aza-[5]triangulene and undoped [5]triangulene. Shown underneath are the accompanying spin-polarized DFT calculations for each molecule.

The calculated energy levels for gas-phase aza-[5]triangulene in its neutral and positively charged states are shown alongside those of undoped [5]triangulene in figure 5, along with the calculated total spins for all three molecules. When aza-[5]triangulene is positively charged, its spin-state is found to be a high-spin quintet (S = 2) with $D_{3h}$ symmetry, an increase of 1/2 over its neutral form. As positively charged aza-[5]triangulene is isoelectronic with undoped [5]triangulene, they both possess the same total spin. However, the general spread of their frontier orbitals differs, particularly in the unoccupied region. Undoped [5]triangulene has a two sets of quasi-degenerate frontier SOMOs and SUMOs, whereas the frontier SUMOs of aza-[5]triangulene in both its forms have more of a spread in energy. The calculated SOMO-SUMO gaps are also significantly different: the calculated energy gap of undoped [5]triangulene is 4.12 eV, compared to 3.07 eV for neutral aza-[5]triangulene and 3.56 eV for the positively charged molecule. The 'real' gap measured for the positively charged aza-[5]triangulene in d$I$/d$V$ experiments on Au(111) is approximately 1.1 eV, significantly lower than the gas-phase DFT. DFT calculations for gas-phase molecules are known to overestimate energy gaps when compared to adsorbed species in experiments,[47] as the substrate and its associated screening/interactions with the molecule are not taken into account.

**Conclusions:**

In conclusion, we have shown *via* spin-polarized theoretical calculations and on-surface synthesis how the symmetry and radical character of centrosymmetric aza-triangulenes is intertwined as their size is increased. Our calculations reveal that these molecules can be separated into two symmetry classes, depending on the location of the central nitrogen atom relative to the two underlying carbon sublattices. We extend the conclusions of previous studies, showing that only the class of aza-triangulenes with a nitrogen located on a different sublattice to the zigzag edge carbons can undergo Jahn-Teller distortions to reduce their symmetry and total spin. As a result of this, all of the centrally-doped aza-triangulenes, regardless of symmetry class, are shown to have one fewer radical than their undoped counterparts.

In addition to this, we have demonstrated the on-surface synthesis of π-extended aza-[5]triangulene, a member of the higher symmetry class, *via* a simple one-step annealing process on Cu(111) and Au(111). In combination with DFT calculations, we have characterized the electronic properties of the molecule on both surfaces, showing how it is most likely positively charged on Au(111) with a high-spin quintet state of S = 2. We believe that this study offers further insight into the relationship between dopant position and the radical character of some types of high spin nanographenes, which may aid their future design and development for applications.

**Methods:**

*Synthetic methods*

The synthesis of precursor **1** is described in detail in the supporting information.

*STM/AFM methods*

All STM and AFM experiments were performed under ultra-high vacuum conditions ($10^{-11}$ to $10^{-10}$ mbar) at low temperatures. All SPM images presented were recorded at 4.3 K. The Cu(111) and Au(111) single crystal substrates were cleaned prior to molecular deposition via cycles of argon sputtering (2 x $10^{-5}$ mbar, 1-2 keV) and annealing (400-500°C). Precursor **1** was sublimed under UHV conditions ($10^{-9}$ mbar regime) at 140°C onto the room temperature substrates. The samples were then post-annealed on the preparation chamber manipulator to 250°C (Cu(111)) and 310°C (Au(111)) for approximately 15 minutes. In the case of Au(111), the sample was placed onto a pre-heated manipulator to rapidly anneal the crystal and prevent desorption of the precursor, which in previous experiments had been found to start occurring at around 200°C.

CO (for tip-functionalization) was deposited onto the substrates at low temperature (4-7 K) via a leak valve mounted on the STM chamber at a pressure of approximately $1 \times 10^{-8}$ mbar. Picking up CO for bond-resolving techniques was typically achieved by scanning areas at higher tunnelling currents (1 nA or more) with negative bias voltages for Au(111) and positive bias voltages for Cu(111). BR-STM and nc-AFM images were recorded with low bias voltages (5 mV) in constant height mode. The QPlus sensor used for nc-AFM imaging had a Q-factor of around 5-10k, with a typical oscillation amplitude setpoint of 50-70 pm.

The built-in Nanonis digital lock-in was utilized for d$I$/d$V$ images and spectra. An oscillation amplitude of 20-30 mV was used for wide-range spectra, with an oscillation of 3 mV for spectra recorded around 0 V that were trying to capture narrow Kondo features.

Some of the images presented in the main text and SI were FFT-filtered to remove high frequency electronic noise that was associated with the QPlus sensor used. SPM images were analyzed using WSxM[48] and Gwyddion.[49]

*Calculation methods*

For details of the calculations, please refer to the theoretical supporting information & figures section.

**Acknowledgements:**
J. Lu acknowledges the support from MOE grants (MOE2019-T2-2-044, MOE T2EP50121-0008, MOE-T2EP10221-0005) and Agency for Science, Technology and Research (A*STAR) under its AME IRG Grant (Project715 No. M21K2c0113). J. Wu acknowledges the financial support from NRF Investigatorship award (NRF-NRFI05-2019-0005). Y. He acknowledges the support from the China Scholarship Council (CSC No. 202208330284). S. Song acknowledges the support from A*STAR under its AME YIRG Grant (M22K3c0094). J. Su acknowledges the support from Agency for Science, Technology and Research (A*STAR) Advanced Manufacturing & Engineering (AME) Young Individual Research Grant (YIRG) A2084c0171. P. Hawrylak, D. Miravet and A. Wania Rodrigues acknowledge support by the Quantum Sensors and Applied Quantum Computing Challenge Programs at the National Research Council of Canada, by NSERC Discovery Grant No. RGPIN- 2019-05714, and University of Ottawa Research Chair in Quantum Theory of Quantum Materials, Nanostructures, and Devices.

**Supporting Information Available:**

Details of precursor synthesis; computational methods; calculated energy levels/orbital densities of all of the aza-[n]triangulenes in this study; calculations for aza-[5]triangulene in different charge states; aza-triangulenes drawn with different resonance structures; table summarizing radicals in different resonance forms; additional STM images; comparison between nc-AFM and particle-probe simulation; hydrogenated aza-[5]triangulene; tip-induced reactions; DFT-optimized adsorption geometries; d$I$/d$V$ of aza-[5]triangulene on Cu(111); comparison of d$I$/d$V$ images with different tips; comparison of orbital densities and d$I$/d$V$ images; spectra of compounds involved in synthesis.

**References:**

(1) Pavliček, N.; Mistry, A.; Majzik, Z.; Moll, N.; Meyer, G.; Fox, D. J.; Gross, L. Synthesis and Characterization of Triangulene. *Nat. Nanotechnol.* **2017**, *12*, 308–311.

(2) Valenta, L.; Juríček, M. The Taming of Clar's Hydrocarbon. *Chem. Commun.* **2022**, *58*, 10896–10906.

(3) Su, J.; Telychko, M.; Song, S.; Lu, J. Triangulenes: From Precursor Design to On-Surface Synthesis and Characterization. *Angew. Chem. Int. Ed.* **2020**, *59*, 7658–7668.

(4) Hawrylak, P.; Peeters, F.; Ensslin, K. Carbononics – Integrating Electronics, Photonics and Spintronics with Graphene Quantum Dots. *Phys. Status Solidi RRL – Rapid Res. Lett.* **2016**, *10*, 11–12.

(5) Güçlü, A. D.; Potasz, P.; Korkusinski, M.; Hawrylak, P. *Graphene Quantum Dots*; NanoScience and Technology; Springer: Berlin, Heidelberg, 2014.

(6) Lieb, E. H. Two Theorems on the Hubbard Model. *Phys. Rev. Lett.* **1989**, *62*, 1201–1204.

(7) Ovchinnikov, A. A. Multiplicity of the Ground State of Large Alternant Organic Molecules with Conjugated Bonds. *Theor. Chim. Acta* **1978**, *47*, 297–304.

(8) Mishra, S.; Beyer, D.; Eimre, K.; Liu, J.; Berger, R.; Gröning, O.; Pignedoli, C. A.; Müllen, K.; Fasel, R.; Feng, X.; Ruffieux, P. Synthesis and Characterization of π-Extended Triangulene. *J. Am. Chem. Soc.* **2019**, *141*, 10621–10625.

(9) Su, J.; Telychko, M.; Hu, P.; Macam, G.; Mutombo, P.; Zhang, H.; Bao, Y.; Cheng, F.; Huang, Z.-Q.; Qiu, Z.; Tan, S. J. R.; Lin, H.; Jelínek, P.; Chuang, F.-C.; Wu, J.; Lu, J. Atomically Precise Bottom-up Synthesis of π-Extended [5]Triangulene. *Sci. Adv.* **2019**, *5*, eaav7717.

(10) Mishra, S.; Xu, K.; Eimre, K.; Komber, H.; Ma, J.; Pignedoli, C. A.; Fasel, R.; Feng, X.; Ruffieux, P. Synthesis and Characterization of [7]Triangulene. *Nanoscale* **2021**, *13*, 1624–1628.

(11) Potasz, P.; Güçlü, A. D.; Hawrylak, P. Zero-Energy States in Triangular and Trapezoidal Graphene Structures. *Phys. Rev. B* **2010**, *81*, 033403.

(12) Wang, T.; Berdonces-Layunta, A.; Friedrich, N.; Vilas-Varela, M.; Calupitan, J. P.; Pascual, J. I.; Peña, D.; Casanova, D.; Corso, M.; de Oteyza, D. G. Aza-Triangulene: On-Surface Synthesis and Electronic and Magnetic Properties. *J. Am. Chem. Soc.* **2022**, *144*, 4522–4529.

(13) Mishra, S.; Beyer, D.; Eimre, K.; Kezilebieke, S.; Berger, R.; Gröning, O.; Pignedoli, C. A.; Müllen, K.; Liljeroth, P.; Ruffieux, P.; Feng, X.; Fasel, R. Topological Frustration Induces Unconventional Magnetism in a Nanographene. *Nat. Nanotechnol.* **2019**, *15*, 22–28.

(14) Mishra, S.; Catarina, G.; Wu, F.; Ortiz, R.; Jacob, D.; Eimre, K.; Ma, J.; Pignedoli, C. A.; Feng, X.; Ruffieux, P.; Fernández-Rossier, J.; Fasel, R. Observation of Fractional Edge Excitations in Nanographene Spin Chains. *Nature* **2021**, *598*, 287–292.

(15) Hieulle, J.; Castro, S.; Friedrich, N.; Vegliante, A.; Lara, F. R.; Sanz, S.; Rey, D.; Corso, M.; Frederiksen, T.; Pascual, J. I.; Peña, D. On-Surface Synthesis and Collective Spin Excitations of a Triangulene-Based Nanostar. *Angew. Chem. Int. Ed.* **2021**, *60*, 25224–25229.

(16) Berdonces-Layunta, A.; Lawrence, J.; Edalatmanesh, S.; Castro-Esteban, J.; Wang, T.; Mohammed, M. S. G.; Colazzo, L.; Peña, D.; Jelínek, P.; De Oteyza, D. G. Chemical Stability of (3,1)-Chiral Graphene Nanoribbons. *ACS Nano* **2021**, *15*, 5610–5617.

(17) Clar, E.; Stewart, D. G. Aromatic Hydrocarbons. LXV. Triangulene Derivatives1. *J. Am. Chem. Soc.* **1953**, *75*, 2667–2672.

(18) Wei, H.; Hou, X.; Xu, T.; Zou, Y.; Li, G.; Wu, S.; Geng, Y.; Wu, J. Solution-Phase Synthesis and Isolation of An Aza-Triangulene and Its Cation in Crystalline Form. *Angew. Chem. Int. Ed.* **2022**, *61*, e202210386.

(19) Valenta, L.; Mayländer, M.; Kappeler, P.; Blacque, O.; Šolomek, T.; Richert, S.; Juríček, M. Trimesityltriangulene: A Persistent Derivative of Clar's Hydrocarbon. *Chem. Commun.* **2022**, *58*, 3019–3022.

(20) Arikawa, S.; Shimizu, A.; Shiomi, D.; Sato, K.; Shintani, R. Synthesis and Isolation of a Kinetically Stabilized Crystalline Triangulene. *J. Am. Chem. Soc.* **2021**, *143*, 19599–19605.

(21) Nakatsuka, S.; Gotoh, H.; Kinoshita, K.; Yasuda, N.; Hatakeyama, T. Divergent Synthesis of Heteroatom-Centered 4,8,12-Triazatriangulenes. *Angew. Chem. Int. Ed.* **2017**, *56*, 5087–5090.

(22) Holt, C. J.; Wentworth, K. J.; Johnson, R. P. A Short and Efficient Synthesis of the [3]Triangulene Ring System. *Angew. Chem. Int. Ed.* **2019**, *58*, 15793–15796.

(23) Ravat, P.; Blacque, O.; Juríček, M. Benzo[Cd]Triangulene: A Spin 1/2 Graphene Fragment. *J. Org. Chem.* **2020**, *85*, 92–100.

(24) Voznyy, O.; Güçlü, A. D.; Potasz, P.; Hawrylak, P. Effect of Edge Reconstruction and Passivation on Zero-Energy States and Magnetism in Triangular Graphene Quantum Dots with Zigzag Edges. *Phys. Rev. B* **2011**, *83*, 165417.

(25) Li, J.; Sanz, S.; Castro-Esteban, J.; Vilas-Varela, M.; Friedrich, N.; Frederiksen, T.; Peña, D.; Pascual, J. I. Uncovering the Triplet Ground State of Triangular Graphene Nanoflakes Engineered with Atomic Precision on a Metal Surface. *Phys. Rev. Lett.* **2020**, *124*, 177201.

(26) Mishra, S.; Beyer, D.; Eimre, K.; Ortiz, R.; Fernández-Rossier, J.; Berger, R.; Gröning, O.; Pignedoli, C. A.; Fasel, R.; Feng, X.; Ruffieux, P. Collective All-Carbon Magnetism in Triangulene Dimers. *Angew. Chem. Int. Ed.* **2020**, *59*, 12041–12047.

(27) Cheng, S.; Xue, Z.; Li, C.; Liu, Y.; Xiang, L.; Ke, Y.; Yan, K.; Wang, S.; Yu, P. On-Surface Synthesis of Triangulene Trimers via Dehydration Reaction. *Nat. Commun.* **2022**, *13*, 1705.

(28) Su, J.; Fan, W.; Mutombo, P.; Peng, X.; Song, S.; Ondráček, M.; Golub, P.; Brabec, J.; Veis, L.; Telychko, M.; Jelínek, P.; Wu, J.; Lu, J. On-Surface Synthesis and Characterization of [7]Triangulene Quantum Ring. *Nano Lett.* **2021**, *21*, 861–867.

(29) Blackwell, R. E.; Zhao, F.; Brooks, E.; Zhu, J.; Piskun, I.; Wang, S.; Delgado, A.; Lee, Y.-L.; Louie, S. G.; Fischer, F. R. Spin Splitting of Dopant Edge State in Magnetic Zigzag Graphene Nanoribbons. *Nature* **2021**, *600*, 647–652.

(30) Wen, E. C. H.; Jacobse, P. H.; Jiang, J.; Wang, Z.; McCurdy, R. D.; Louie, S. G.; Crommie, M. F.; Fischer, F. R. Magnetic Interactions in Substitutional Core-Doped Graphene Nanoribbons. *J. Am. Chem. Soc.* **2022**, *144*, 13696–13703.

(31) Zhu, G.; Jiang, Y.; Wang, Y.; Wang, B.-X.; Zheng, Y.; Liu, Y.; Kang, L.-X.; Li, Z.; Guan, D.; Li, Y.; Zheng, H.; Liu, C.; Jia, J.; Lin, T.; Liu, P.-N.; Li, D.-Y.; Wang, S. Collective Quantum Magnetism in Nitrogen-Doped Nanographenes. *J. Am. Chem. Soc.* **2023**, *145*, 7136–7146.

(32) Li, Z.; Tang, Y.; Guo, J.; Zhang, J.; Deng, M.; Xiao, W.; Li, F.; Yao, Y.; Xie, S.; Yang, K.; Zeng, Z. Stair-like Narrow N-Doped Nanographene with Unusual Diradical Character at the Topological Interface. *Chem* **2023**, *9*, 1281–1294.

(33) Wang, X.-Y.; Yao, X.; Narita, A.; Müllen, K. Heteroatom-Doped Nanographenes with Structural Precision. *Acc. Chem. Res.* **2019**, *52*, 2491–2505.

(34) Kang, D.; Zhang, S.; Ju, W.; Zuo, Z.; Wang, Z. Heteroatom-Doped Clar's Goblet: Tunable Magnetic Order and Programmable Spin Logic Gate. *Appl. Phys. Lett.* **2021**, *119*, 192408.

(35) Kervella, Y.; Andrés Castán, J. M.; Avalos-Quiroz, Y. A.; Khodr, A.; Eynaud, Q.; Koganezawa, T.; Yoshimoto, N.; Margeat, O.; Rivaton, A.; Riquelme, A. J.; Mwalukuku, V. M.; Pécaut, J.; Grévin, B.; Videlot-Ackermann, C.; Ackermann, J.; Demadrille, R.; Aumaître, C. Star-Shape Non-Fullerene Acceptor Featuring an Aza-Triangulene Core for Organic Solar Cells. *J. Mater. Chem. C* **2023**, *11*, 8161–8169.

(36) Arikawa, S.; Shimizu, A.; Shiomi, D.; Sato, K.; Takui, T.; Sotome, H.; Miyasaka, H.; Murai, M.; Yamaguchi, S.; Shintani, R. A Kinetically Stabilized Nitrogen-Doped Triangulene Cation: Stable and NIR Fluorescent Diradical Cation with Triplet Ground State. *Angew. Chem. Int. Ed.* n/a (n/a), e202302714.

(37) Zhao, M.; Miao, Q. Design, Synthesis and Hydrogen Bonding of B3N6-[4]Triangulene. *Angew. Chem.* **2021**, *133*, 21459–21464.

(38) Chen, C.; Lu, J.; Lv, Y.; Yan, Y.; Sun, Q.; Narita, A.; Müllen, K.; Wang, X.-Y. Heteroatom-Edged [4]Triangulene: Facile Synthesis and Two-Dimensional On-Surface Self-Assemblies. *Angew. Chem. Int. Ed.* **2022**, *61*, e202212594.

(39) Chen, X.; Tan, D.; Dong, J.; Ma, T.; Duan, Y.; Yang, D.-T. [4]Triangulenes Modified by Three Oxygen-Boron-Oxygen (OBO) Units: Synthesis, Characterizations, and Anti-Kasha Emissions. *J. Phys. Chem. Lett.* **2022**, *13*, 10085–10091.

(40) Jahn, H. A.; Bragg, W. H. Stability of Polyatomic Molecules in Degenerate Electronic States II-Spin Degeneracy. *Proc. R. Soc. Lond. Ser. - Math. Phys. Sci.* **1938**, *164*, 117–131.

(41) Lawrence, J.; Sosso, G. C.; Đorđević, L.; Pinfold, H.; Bonifazi, D.; Costantini, G. Combining High-Resolution Scanning Tunnelling Microscopy and First-Principles Simulations to Identify Halogen Bonding. *Nat Commun* **2020**, *11*, 2103.

(42) Kawai, S.; Nakatsuka, S.; Hatakeyama, T.; Pawlak, R.; Meier, T.; Tracey, J.; Meyer, E.; Foster, A. S. Multiple Heteroatom Substitution to Graphene Nanoribbon. *Sci. Adv.* **2018**, *4*, eaar7181.

(43) Hapala, P.; Kichin, G.; Wagner, C.; Tautz, F. S.; Temirov, R.; Jelínek, P. Mechanism of High-Resolution STM/AFM Imaging with Functionalized Tips. *Phys. Rev. B* **2014**, *90*, 085421.

(44) Hapala, P.; Temirov, R.; Tautz, F. S.; Jelínek, P. Origin of High-Resolution IETS-STM Images of Organic Molecules with Functionalized Tips. *Phys. Rev. Lett.* **2014**, *113*, 226101.

(45) Lawrence, J.; Berdonces-Layunta, A.; Edalatmanesh, S.; Castro-Esteban, J.; Wang, T.; Jimenez-Martin, A.; de la Torre, B.; Castrillo-Bodero, R.; Angulo-Portugal, P.; Mohammed, M. S. G.; Matěj, A.; Vilas-Varela, M.; Schiller, F.; Corso, M.; Jelinek, P.; Peña, D.; de Oteyza, D. G. Circumventing the Stability Problems of Graphene Nanoribbon Zigzag Edges. *Nat. Chem.* **2022**, *14*, 1451–1458.

(46) Ruffieux, P.; Wang, S.; Yang, B.; Sanchez-Sanchez, C.; Liu, J.; Dienel, T.; Talirz, L.; Shinde, P.; Pignedoli, C. A.; Passerone, D.; Dumslaff, T.; Feng, X.; Müllen, K.; Fasel, R. On-Surface Synthesis of Graphene Nanoribbons with Zigzag Edge Topology. *Nature* **2016**, *531*, 489–492.

(47) Goiri, E.; Borghetti, P.; El-Sayed, A.; Ortega, J. E.; De Oteyza, D. G. Multi-Component Organic Layers on Metal Substrates. *Adv. Mater.* **2016**, *28*, 1340–1368.

(48) Horcas, I.; Fernández, R.; Gómez-Rodríguez, J. M.; Colchero, J.; Gómez-Herrero, J.; Baro, a M. WSXM: A Software for Scanning Probe Microscopy and a Tool for Nanotechnology. *Rev. Sci. Instrum.* **2007**, *78*, 013705.

(49) Nečas, D.; Klapetek, P. Gwyddion: An Open-Source Software for SPM Data Analysis. *Cent. Eur. J. Phys.* **2012**, *10*, 181–188.